\documentclass[preprint,aps]{revtex4}

\usepackage{graphicx}
\usepackage{dcolumn}
\usepackage{bm}
\usepackage{natbib}
\begin{document}

\preprint{APS/123-QED}

\title{Abrupt transition in quasiparticle dynamics at optimal doping in
a cuprate superconductor system}
\author{N. Gedik}
\altaffiliation[Present address: ]{Laboratory for Molecular Sciences, Arthur Amos Noyes
Laboratory of Chemical Physics, California Institute of Technology, Pasadena, CA 91125 }
\email{gedik@caltech.edu}
\author{J. Orenstein}
\affiliation{Physics Department, University of California,
Berkeley and \\ Materials Science Division, Lawrence Berkeley
National Laboratory, Berkeley, CA 94720}
\author{S. Ono}\author{Yasushi Abe}\altaffiliation[Present address: ]{CERC, AIST, Tsukuba 205-8562, Japan}
\author{Yoichi Ando}
\affiliation{Central Research Institute of Electric Power
Industry, Komae, Tokyo 201-8511, Japan}

\begin{abstract}
We report time-resolved measurements of the photoinduced change in
reflectivity, $\Delta R$, in the
Bi$_2$Sr$_2$Ca$_{1-y}$Dy$_y$Cu$_2$O$_{8+\delta}$ (BSCCO) system of
cuprate superconductors as a function of hole concentration. We
find that the kinetics of quasiparticle decay and the sign of
$\Delta R$ both change abruptly where the superconducting
transition temperature $T_c$ is maximal. These coincident changes
suggest that a sharp transition in quasiparticle dynamics takes
place precisely at optimal doping in the BSCCO system.
\end{abstract}
\maketitle


Pump and probe methods in optical spectroscopy have opened a new
window on the properties of quasiparticles in cuprate
superconductors and other highly correlated electron systems
\cite{averitt2002}. In experiments based on these methods,
ultra-short pump pulses inject quasiparticles at densities that
are continuously variable from well above to well below the
thermal equilibrium level. Time-delayed probe pulses measure
changes in the reflectivity or transmissivity that result from the
presence of nonequilibrium quasiparticles, providing information
about their recombination rates, transport, and optical
properties. These studies have been carried out extensively in the
cuprate superconductors, yielding a rich, complex, yet poorly
understood array of experimental observations. One of the central
observations, and possibly the most puzzling, has been the
behavior of the quasiparticle recombination rate, $\gamma$, as a
function of temperature, $T$, and photoinjected density, $\Delta
n_{ph}$.  Two classes of behavior are found: in class (1) $\gamma$
appears to vanish as $T$
\cite{gay1999,smith1999,demsar2001,segre2002,schneider2002} and
$\Delta n_{ph}$ \cite{gay1999, segre2002, gedik2003, gedik2004}
tend to zero, while in class (2) $\gamma$ remains essentially
constant with decreasing $T$ \cite{han1990, averitt2001,
demsar1999} and $\Delta n_{ph}$ \cite{albrecht1993}. Another,
seemingly distinct, puzzle concerns the sign of the photoinduced
change in sample reflectivity, $\Delta R$, which can be either
positive or negative
\cite{han1990,eesley1990,reitze1992,gay1999,dvorsek2002}.

Here we report measurements of $\Delta R$ and $\gamma$ in the
Bi$_2$Sr$_2$Ca$_{1-y}$Dy$_y$Cu$_2$O$_{8+\delta}$ (BSCCO) system of
cuprate superconductors as a function of hole concentration, $x$,
that considerably clarify the conditions under which these
behaviors appear. As discussed below, the key to successfully
exploring the BSCCO system was to eliminate the effects of
laser-induced heating. Once this is accomplished, we find that the
dynamics change from class (1) to (2) at exactly $x_m$, the value
for which the superconducting transition temperature $T_c$ is
maximal. Moreover, we find that the sign of $\Delta R$ reverses at
$x_m$ as well. These coincident changes suggest that an abrupt
transition in quasiparticle dynamics takes place precisely at
optimal doping in the BSCCO system.

Time-resolved optical spectroscopy was performed using pump and
probe pulses of photon energy 1.5 eV and duration 80 fs from a
mode-locked Ti:Sapphire oscillator.  Because the BSCCO crystals
are optically thick at the laser wavelength of 820 nm, the changes
in optical response were probed by measuring the reflected probe
power.  Fig. 1 is a plot of the initial reflectivity change,
$\Delta R$, normalized to the reflectivity $R$, as a function of
the energy per area, $\Phi_L$, deposited by each pump pulse. The
underdoped ($T_c$=71 K) sample was thermally anchored to a Cu
plate maintained at 5 K. The $\Delta R$ values plotted as open
symbols were measured using the full repetition rate of the
oscillator, which is 90 MHz. At this repetition rate, the slope of
$\Delta R/R$ vs. laser intensity changes abruptly when $\Phi_L$
reachess 0.8 $\mu$J/cm$^2$. This effect was not observed in
studies of underdoped YBa$_2$Cu$_3$O$_{6.5}$ crystals with similar
values of $T_c$ and the same conditions of photoexcitation
\cite{segre2002,gedik2004}.

\begin{figure}
\includegraphics[width=3.5in]{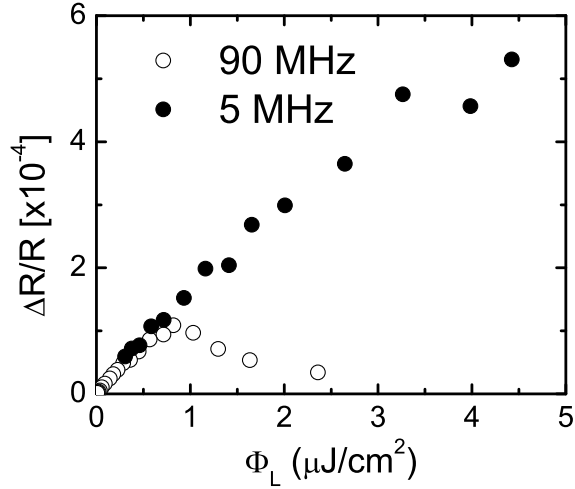}
\caption{Fractional change in reflectivity at 1.5 eV as a function
of pump energy/(area-pulse), for two values of the pump pulse
repetition rate.} \label{fig:first}
\end{figure}

The decrease in $\Delta R/R$ with increasing $\Phi_L$ occurs when
photoexcitation begins to drive the surface of the sample into the
normal state. The origin of this effect is the steady state
increase in the surface temperature of the sample due to
laser-induced heating. The effects of surface heating are more
severe than in the YBCO system because c-axis thermal conductivity
of BSCCO is much smaller \cite{crommie1991}. To overcome the
laser-heating problem, we inserted an acousto-optic pulse-picker
at the laser output. The solid symbols in Fig. 1 indicate values
of $\Delta R/R$ obtained when the pulse-picker reduces the 90 MHz
pulse repetition rate of the laser to 5 MHz (at the same time the
diameter of the illuminated area on the sample was reduced from 75
to 30 $\mu$). When the average power is thus reduced (by a factor
of $\sim$ 100) the discontinuity in the slope of $\Delta R$
\textit{vs.} $\Phi_L$ disappears and the growth of $\Delta R/R$
with $\Phi_L$ is essentially linear. Further reduction of the
repetition rate to 2.5 MHz produced no further changes in the
either the amplitude or the subsequent decay of $\Delta R/R$,
indicating that effects of laser heating are negligible at this
power level.

Eliminating problems associated with laser heating makes it
possible to investigate the nonequilibrium state of the BSCCO
family of superconductors at low temperature.  In this work we
studied eight BSCCO crystals whose $T_c$'s range from 42 K
(underdoped) to 77 K (overdoped).  To control hole concentration
over a wide range, we combined both oxygen tuning and Dy-doping.
We achieved the target $T_c$'s by floating-zone growth of three
different kinds of single crystals with the following Bi:Sr:Ca:Dy
ratio : (a) 2.10 : 1.91 : 1.03 : 0, (b) 2.11 : 1.88 : 0.79 : 0.23,
and (c) 2.13 : 1.82 : 0.70 : 0.36. The crystals were subsequently
annealed for 2 to 14 days, depending on the annealing temperature,
in air or an argon+oxygen environment.

We now describe the evolution of the photoinduced $\Delta R$ with
hole concentration. Fig. 2 presents an overview of the changes,
showing $\Delta R$ (at $\Phi_L=0.9$ $\mu$J/cm$^2$) as a function
of time for three representative samples: underdoped ($T_c$=71 K),
optimally-doped ($T_c$=94.5 K), and overdoped ($T_c$=77 K).
$\Delta R$ changes from positive for the underdoped sample to
negative for the overdoped sample. In the sample with $T_c$=94.5
K, $\Delta R$ is a superposition of signals of both sign,
indicating that the crossover takes place precisely at optimal
doping.

\begin{figure}
\includegraphics[width=3.5in]{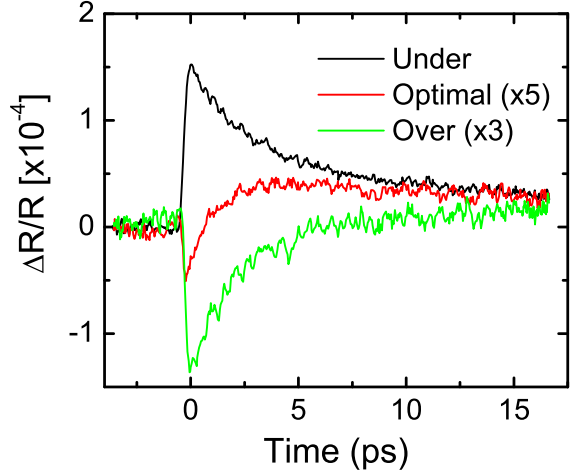}
\caption{Fractional change in reflectivity as a function of time
following pulsed photoexcitation, comparing the response of under,
optimal, and overdoped samples.} \label{fig:second}
\end{figure}

Figs. 3(a) and 3(b) illustrate the crossover in kinetics that
takes place at the same hole concentration at which $\Delta R$
changes sign. The top panels show $\Delta R(t)/R$ at different
values of $\Phi_L$, for the same under and overdoped samples as in
Fig. 2. In order to compare the decay rates at different $\Phi_L$,
we scale the amplitude to the same value near time zero. For each
set of curves, a plot of the initial decay rate, $\gamma(0)$, vs.
$\Phi_L$ appears below. In underdoped samples
$\gamma(0)\propto\Phi_L$, with essentially zero intercept (we
estimate an experimental uncertainty in $\gamma(0)$ of 0.005
ps$^{-1}$, approximately $10^{-2}$ of the maximum $\gamma(0)$ of
0.6 ps$^{-1}$). The adjacent panels illustrate that the kinetics
of photoexcitations in overdoped samples are substantially
different - the decay rate of the excited state remains large as
$\Phi_L$ is lowered.

\begin{figure}
\includegraphics[width=5in]{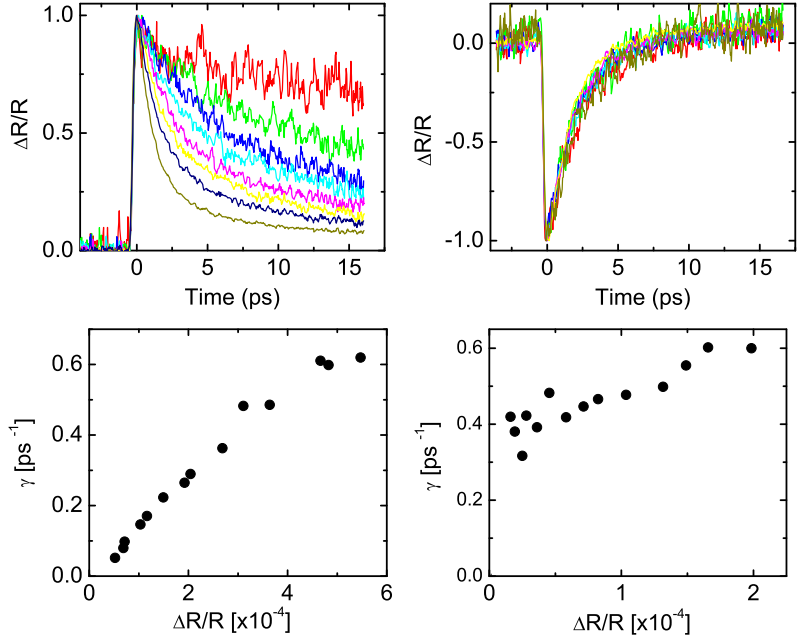}
\caption{Top panels: Fractional change in reflectivity as a
function of time for an \textbf{(a)} underdoped ($T_c$=71 K and
\textbf{(b)} overdoped sample (T$_c$=77 K), for several values of
the pump fluence in the range from 0.1 to 1.0 $\mu$J/cm$^2$. The
plots are scaled to have the same value at $t=0$, illustrating
that the decay rate depends on fluence in underdoped samples but
not in overdoped samples.  Bottom panels \textbf{(c)} and
\textbf{(d)}: Initial decay rate as a function of initial $\Delta
R/R$ for the curves directly above.} \label{fig:third}
\end{figure}

The abruptness of the change in sign and decay rate of $\Delta R$
with hole concentration, $x$, is shown Figs. 4(a) and 4(b). Fig.
4(a) is a plot of $\Delta R$ vs. $x$ measured at 0.2 ps after
arrival of the pump pulse. The values of $x$ were obtained from
the $T_c$'s using the empirical formula given in Ref.
\cite{presland1991}. Fig. 4(b) shows the initial decay rate of
$\Delta R$ for the same set of samples, measured at $\Phi_L=0.3$
$\mu$J/cm$^2$. At this relatively low $\Phi_L$, the lifetime
changes by a factor $\sim$ 50 as $x$ varies from just below to
just above $x_m$. In the following, we discuss the origin of the
two transitions, starting with the sign of $\Delta R$.

\begin{figure}
\includegraphics[width=3in]{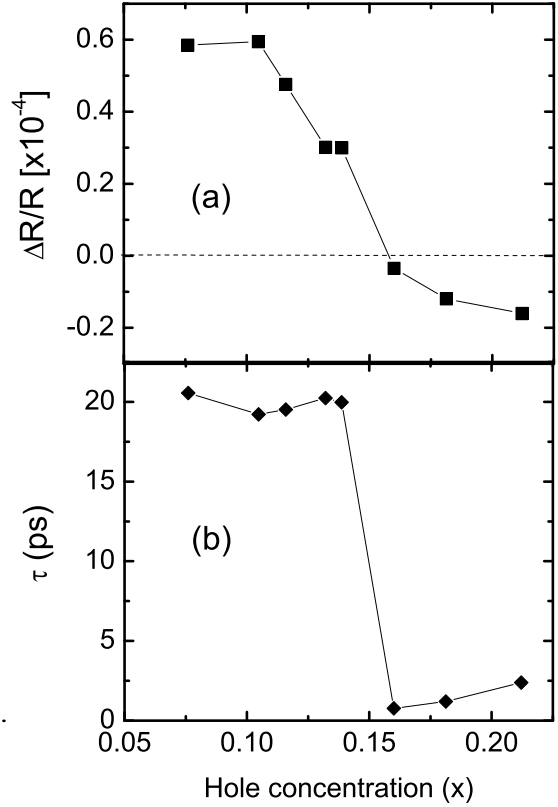}
\caption{Fractional change in reflectivity (top panel) and initial
lifetime (bottom panel) as a function of hole concentration. All
data measured at pump laser fluence 0.3 $\mu$J/cm$^2$. The
collapse of the quasiparticle lifetime and sign change of $\Delta
R$ both occur at optimal doping.} \label{fig:fourth}
\end{figure}

To analyze the sign change of $\Delta R$ we consider the change in
the real part of the dielectric function, $\Delta\epsilon_1$,
caused by the presence of nonequilibrium excitations. In general,
a change in the occupation of states shifts optical spectral
weight from one frequency range to another, conserving the total
weight. The sign of $\Delta\epsilon_1$ caused by such shifts
depends on the relative ordering of the measurement frequency,
$\omega_0$, and frequencies where spectral weight is lost and
gained.  For a superconductor described by BCS theory, an increase
in the number density of quasiparticles shifts spectral weight
from the condensate $\delta$-function at $\omega=0$ to frequencies
near the quasiparticle scattering rate, $1/\tau$, or the gap
$\Delta$, in the clean, or dirty limit, respectively. Assuming
that $\omega_0\gg\max\{\Delta, 1/\tau\}$, the Kramers-Kronig
relations imply that $\Delta\epsilon_1\simeq-(8/\omega_0^4)\int
d\omega\Delta\sigma_1(\omega)\omega^2$, where $\Delta\sigma_1$ is
spectrum of the conductivity that was removed from the
$\delta$-function \cite{segre2002}.  As $\Delta\sigma_1$ is
positive, the resulting $\Delta\epsilon_1$ must be negative. From
the Fresnel equation and literature values \cite{hwang2004} for
$\epsilon(\omega)$ we obtain $R^{-1}\partial R/\partial
\epsilon_1=0.02$ at the measurement energy of 1.5 eV. Thus the
spectral weight shifts expected for a BCS superconductor
correspond to $\Delta R<0$ and can account for the photoinduced
response in overdoped, but not underdoped BSCCO \cite{demsar2003,
gedik2003a}.

To obtain $\Delta\epsilon_1$ of the opposite sign requires that a
fraction of the spectral weight removed from the $\delta$-function
either shifts to $\omega\gg\omega_0$, or becomes broad on the
scale of $\omega_0$. In other words, some spectral weight must be
distributed on the scale of the electronic bandwidth, rather than
the frequency scale of $1/\tau$ or $\Delta$.  In this case
$\Delta\epsilon_1$ acquires a positive contribution, given
approximately by $8A/\omega_0^2$, where $A$ is the portion of the
spectral weight removed from the condensate that is shifted to
high frequencies \cite{molegraaf2002}.  Using this relation, we
can infer the value of $A$ that generates a given positive value
of $\Delta R$. To express $A$ in a fashion that can be compared
with other experiments, we convert conductivity spectral weight to
the electron kinetic energy \cite{kubo1957}, \textit{i.e.}
$A\equiv\int \sigma_1d\omega=(\pi e^2/2\hbar^2
d)\langle-T\rangle$.  The largest $\Delta R$ that we obtain,
$\sim4\times10^{-4}$, corresponds to $\langle -\Delta T \rangle=3$
meV.  We note a possible connection between the photoinduced
$\Delta\epsilon_1$ and the thermally induced $\Delta\epsilon_1$
recently reported in BSCCO \cite{molegraaf2002}. When spectral
weight is removed from the $\delta$-function by raising the
temperature, $\Delta \epsilon_1$ in the near-infrared is positive
and similar in magnitude to the photoinduced changes reported
here.

We next discuss the discontinuous change in decay rate that takes
place at the same hole concentration at which $\Delta R$ changes
sign. The sudden change in rate at $x_m$ marks the transition from
$\gamma(0)\propto\Phi_L$ on the underdoped side of the phase
diagram and to $\gamma(0)$ independent of $\Phi_L$ on the
overdoped side. For underdoped samples, the decay rate is a linear
function of the density of photoinduced excitations. Such
"second-order kinetics" occurs when individual, isolated
excitations are stable (or metastable) and the rate of decay is
limited by the frequency of two-quasiparticle encounters. The
transition to a $\Phi_L$-independent decay rate at optimal doping
indicates that excitations that were metastable on the underdoped
side of the phase diagram become unstable on the overdoped side.
Below, we discuss the nature of these excitations and potential
explanations for the metastable-unstable transition.

The linear dependence of $\Delta R$ on $\Phi_L$ suggests that the
excitations that give rise to $\Delta R$ are not nodal
quasiparticles.  More precisely, it suggests that photoexcited
quasiparticles do not immediately thermalize towards the nodes,
establishing a degenerate Fermi-Dirac distribution with an
effective chemical potential or temperature. If this were the
case, the spectral weight removed from the condensate would be
proportional to $\Delta n_{ph}^{1/2}$ and $\Delta R$ would be
proportional to $\Phi_L^{1/3}$ \cite{schachinger2004}. The
experimental observation that instead $\Delta R\propto\Phi_L$
suggests that photoexcited quasiparticles form a nondegenerate
gas, as do quasiparticles injected in an s-wave superconductor.
The scenario that emerges is that injected quasiparticles cascade
to the antinodal regions of the Brillouin zone where the density
of states is very large. In underdoped samples they remain there,
prevented from thermalizing further (for reasons we speculate on
below) until an encounter with another quasiparticle. The rapid,
$\Delta n_{ph}$-independent decay of $\Delta R$ that appears at
optimal doping may indicate that the bottleneck for thermalization
of antinodal particles disappears above $x_m$.

The scenario described above bears a certain relationship to ARPES
investigations \cite{damascelli2003} of the quasiparticle
self-energy as a function of $x$, particularly to the recent
emphasis on a "dichotomy" between nodal and antinodal excitations
\cite{yoshida2003,zhou2004,kshen2005}.  The dichotomy exists on
the underdoped side, where excitations near the node are coherent
quasiparticles and antinodal excitations are incoherent. With
increasing $x$ the quasiparticle coherence extends further from
the node and appears to encompass the entire Fermi surface in
overdoped samples.  We speculate that the bottleneck in underdoped
materials exists because the incoherent nodal excitations cannot
readily convert to coherent nodal excitations. When the entire FS
becomes coherent the bottleneck is removed.

Further evidence that that the excitations in underdoped samples
are incoherent comes from the second-order recombination
coefficient, $\beta$, defined such that $\gamma(0)=\beta\Delta
n_{ph}$ \cite{rothwarf1967}. To estimate $\beta$ we assume that
each photon creates $\hbar\omega_0/\Delta$ particles, where
$\Delta$ is the antinodal creation energy of $\sim$35 meV.
Converting photon to quasiparticle number as above, we obtain
$\beta\approx 0.1$ cm$^2$/s for all the underdoped samples
measured in this study. If a particle is incoherent, its bandlike
motion is frustrated and the rate of recombination can become
limited by the time needed to diffuse the average distance between
them. For a 2D "diffusion-limited reaction" process $\beta\simeq
D$. It is therefore interesting to compare $\beta$ with the
quantum diffusion of a fermion in 2D. At the localization limit
$k_Fl=1$, $D=1/hN_F$ or $\hbar/2m^\ast$, where $N_F$ and $m^\ast$
are the Fermi level density of states and effective mass,
respectively. If we take $m^\ast=3m$ as suggested from optical
measurements \cite{basov2004}, then $D=0.15$ cm$^2$/s, which is
remarkably close the experimental estimate for $\beta$.

In conclusion, we have observed an abrupt change in quasiparticle
decay rate and the sign of the photoinduced $\Delta R$ in BSCCO
superconductors precisely at optimal doping.  On the underdoped
side, the photoinjected quasiparticles appear to propagate
incoherently and cause some condensate spectral weight to shift to
very high frequencies. The sign of spectral weight shift is
consistent with recent suggestions of a kinetic-energy driven
transition to the superconducting state \cite{molegraaf2002}. The
sudden change in the quasiparticle dynamics at $x_m$ may signal
the onset of antinodal quasiparticle coherence in optimal and
overdoped samples.  The change in the sign of $\Delta R$ at $x_m$
suggests that reduction in kinetic energy with the onset of
superconductivity occurs primarily in the underdoped regime.

This work was supported by DOE-DE-AC03-76SF00098.

\renewcommand{\baselinestretch}{1} \small\normalsize
\bibliography{bscco}
\end{document}